**Graphical Abstract**

# A Theoretical Study on Band-Gap Engineering of $CsCaI_3$ by Si Doping for Photo Voltaic Applications

Krishnaraj Kundavu[1], Parveen Kumar[2], R P Chauhan[1]*

[1]Department of Physics, National Institute of Technology, Kurukshetra 136119, India

[2]Bio-Nanotechnology Lab, CSIR-CSIO, Chandigarh, India

The effect of Silicon doping on $CsCaI_3$ was studied using First Principles calculations. The structural, electronic and optical properties were analyzed. Calculations showed that with increasing Si doping percentage, the band gap decreases and optical absorption increases making it a probable candidate for solar cell applications.

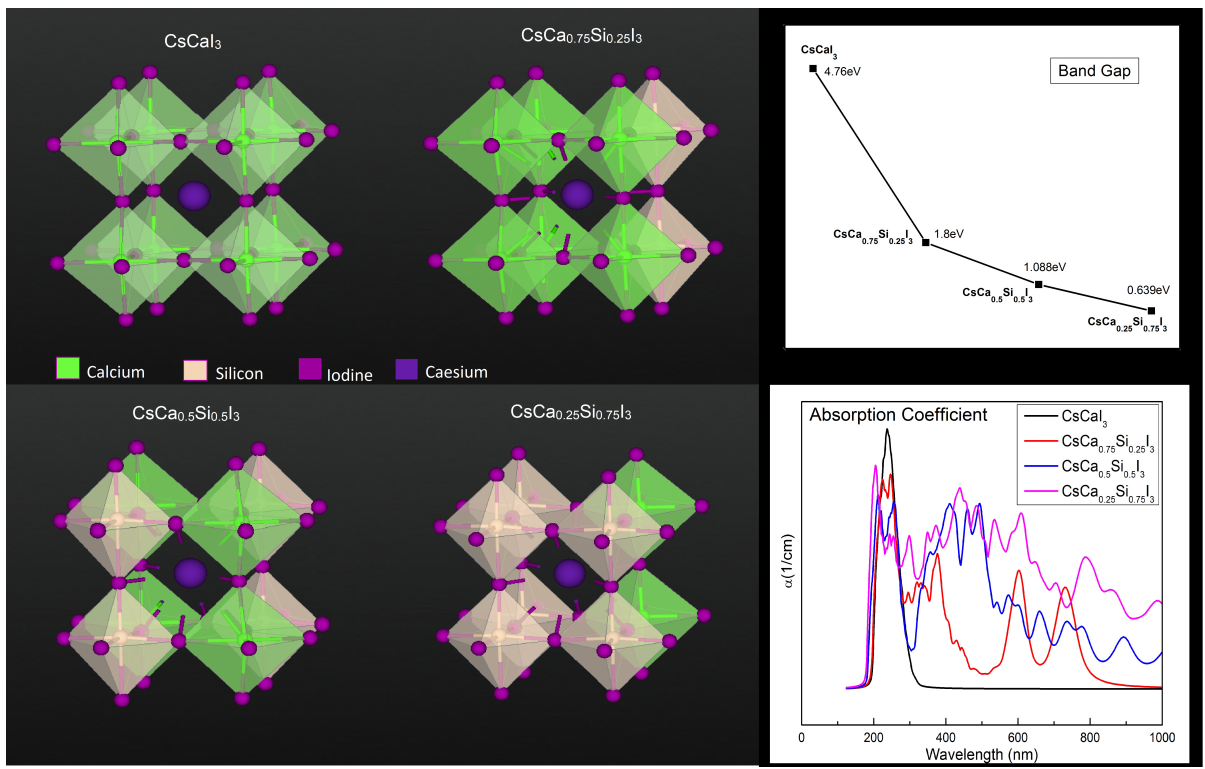

# Title page:

# A Theoretical Study on Band-Gap Engineering of $CsCaI_3$ by Si Doping for Photo Voltaic Applications

Krishnaraj Kundavu[1], Parveen Kumar[2], R P Chauhan[1*]

[1]*Department of Physics, National Institute of Technology, Kurukshetra 136119, India*

[2]*Bio-Nanotechnology Lab, CSIR-CSIO, Chandigarh, India*

Corresponding author: Krishnaraj Kundavu*

Mailing address: Department of Physics, NIT Kurukshetra, Thanesar, Haryana - 136119, India

Tel: +917259742731

Email: krishnaraj868@gmail.com

**Abstract** - Density functional theory based First Principles calculations were used to study the effect of Silicon (Si) doping on the structural, electronic and optical properties of $CsCaI_3$. From our calculations, we predict that $CsCaI_3$ can form stable perovskite structure. It's also observed that after substitutional doping of Si in $CsCaI_3$, the material still can stay in perovskite form. Band structure studies showed that with Si doping, the band gap can be varied from 4.76eV for un-doped $CsCaI_3$ to 0.639eV for 75% Si doped $CsCaI_3$. The optical absorption spectra of the materials showed that Si doping can induce light absorption in the visible region in $CsCaI_3$. Electron Localization Function (ELF) and effective mass calculations show that Si doping can improve the electronic conduction in this material to meet the requirements of photo Voltaic applications.

**1. Introduction**

Metal Halide perovskites ($AMX_3$), where A is either a metal or an organic molecule, M is a di-valent group IV element and X is a halogen atom (Cl, Br and I) have become a well-studied material for solar cell applications as solar absorbents. With extensive research, the efficiency of these materials has gone up from 3.8% [1] to 22.1% [2] in a span of just ten years. Though the cell areas of these solar cells are small (tenths of $cm^2$) [3], encouraging enough efficiency - greater than 15% [4] has also been obtained for large area solar cells with cell area greater than 1 $cm^2$. These perovskites contain earth-abundant essential elements, making them highly promising for low-cost and large-scale Photo Voltaic (PV) applications. The main interest on these perovskite solar cells is because of the solution processability of these materials [5-9].

The flag bearer of this class of materials is Methylammonium Lead Iodide ($MAPbI_3$) which has drawn huge attention in the research community. This materials is found to have sharp and high absorption peak almost 25 times higher than Si and better than GaAs [10-12]. They have long lived photo generated excitons with long diffusion lengths [13-16] and low exciton binding energies. Both electrons and holes have low effective masses causing ambipolar conduction of photo current [17]. In $MAPbI_3$, dissipationless absorption and emission of photons were observed, which enable photons to recycle, leading to utilization of photons within the active layer [18].

Even though these materials have shown impressive properties and scope for utilization in solar cells, there are some challenges stopping their large-scale adoption. The environmental stability of these materials are not good enough to be used in outdoor conditions and also at high temperature [19]. These materials are thermodynamically unstable [20, 21] and it has been observed that in the presence of moisture, these perovskites degrade into Lead halides causing Lead contamination.

Some key properties affecting PV performance need to be optimized, such as reducing the optical band gap of $[CH_3NH_3]PbI_3$ (1.5 eV) [22,23] and $[CH_3NH_3]PbBr_3$ (2.35 eV) [24] to the optimal value of 1.34 eV from the Shockley-Queisser limit [25]. Further enhancement in light capture efficiency can be contributed by organic molecules in some way to absorption in the solar range, and tuning material parameters to eliminate anomalous hysteresis in the current-voltage curves [26], etc.

Many studies have been conducted to tune the absorption efficiency and stability of these materials by compositional engineering [27-31]. The Methyl Ammonium (MA) cation has been replaced by another organic cation formamidinium (FA), Rubidium ($Rb^+$), and even double and triple cation replacements such as MA/FA, FA/Cs and Cs/FA/MA etc. Different combinations of halide anions ($Cl^-$, $Br^-$, $I^-$) were also studied to tune the properties of these perovskites. Recently more importance is given to replace the toxic Lead in these materials with non-toxic, and easily available elements. Some single element replacements that have caught attention are Tin (Sn) [32] and Germanium (Ge) [33]. Even though these replacements show promising properties, they still lack the stability and efficiency to replace Pb based perovskites.

Theoretical knowledge is also necessary along with experimental results to better understand the trends and properties of materials. Trial and error methods have to be replaced with fundamental understanding of the materials to design materials for solar applications. *Ab-initio* calculations help identify various materials suitable as solar absorbents and understand their behavior. Recently many researchers have tried both mono and binary replacements of Pb by systematically analyzing all the possible cations. One *ab-initio* study has shown that Ca/Si and Zn/Si replacements can give high device absorption efficiencies [34]. Extensive studies have been conducted to identify the suitable replacements for Pb in the periodic table through First Principles calculations [35-45]. These studies have created a list of all suitable perovskites for solar cell applications.

In this work, we conducted First Principles calculations based study on band gap engineering of CsCaI$_3$ by doping Si and studying its effect on the structural, electronic and optical properties of the material. The study is intended to overcome the instability caused by the organic cation and the toxicity of lead in the most commonly used Perovskite solar cells (PSCs).

## 2. Methods

For this study Density Functional Theory (DFT) based calculations, implemented as part of Atomistix ToolKit (ATK) package provided by QuantumWise [46] was used. To start with, Crystal Structure Prediction tool available in ATK has been used to predict the structure of CsCaI$_3$ through genetic algorithm. Genetic algorithm solves both constrained and unconstrained optimization problems by mimicking a natural selection process. The algorithm repeatedly modifies/refines a population of individual solutions. The algorithm starts from an initial population of crystals (a generation), and iteratively refines the generation by producing new generations through application of genetic operators such as promotion, heredity, mutation and permutation to the previous generation. A 2x2x2 Monkhorst–Pack k-point grid [47] is used to sample the Brillouin zone to balance between accuracy and computational cost. Four molecules of CsCaI$_3$ were used for the prediction. The total number of generations was set to 10 with 10 individuals in each generation. Force cutoff of 0.01 eV/Å is used for convergence. Number of individuals to be promoted to the next generation was set to 4.

Substitutional doping was represented by replacing the Ca atoms in the 4 molecule unit cell of CsCaI$_3$ by Si atoms. The structural optimization was done by Geometry optimization tool with 3x2x2 Monkhorst–Pack k-point grid sampling. For both Crystal structure prediction and geometry optimization, Generalized Gradient Approximation (GGA) with Perdew-Burke-Ernzerhof (PBE) [48] exchange correlation functional was used.

The band structure and carrier effective masses of the materials were calculated using different exchange correlation functions such as GGA-PBE, DFT-1/2 [50], Hubbard U [53-55], Meta-GGA [56] and Spin Orbit Coupling (SOC) [57]. The results from the above methods are compared to analyze the accuracy of our calculations. Norm conserving SG15 pseudo potentials [58] were used with GGA-PBE and MGGA methods, while fully relativistic SG15-SO pseudo potentials were used for SOC calculations. All the calculations for the properties analysis were done with 12x8x6 Monkhorst–Pack k-point grid sampling.

## 3. Results and Discussion

In general, the first step to study any $MAPbI_3$ like material is by starting with the structure of $MAPbI_3$ and substituting the cations or anions with suitable replacements. Then structures are optimized to get the new structure. Even though this approach is proven to give good results, it is limited by the computational cost of extensive geometry optimization calculations to explore possible structures. In our study, we rather started with the crystal structure prediction functionality provided by ATK package of Quantumwise to obtain the possible structures of $CsCaI_3$. This function uses the genetic algorithms to obtain least energy structures for the given chemical formula. We obtained ten least energy structures containing four molecules of $CsCaI_3$. Out of ten structures obtained, top two structures caught our attention due to the presence of perovskite like cage structure. Figure 1 shows the unit cells of the top 2 structures obtained from crystal structure prediction. Both these structures were similar with only a small change in total energy and cell parameters. The obtained structures had less symmetry compared to the reported orthorhombic structure in literature [59-61]. This can be attributed to the high temperature synthesis process which is known to produce higher symmetry structures. For example, $MAPbI_3$ stabilizes to cubic perovskite structure at elevated temperatures as compared to the tetragonal structure at room temperature.

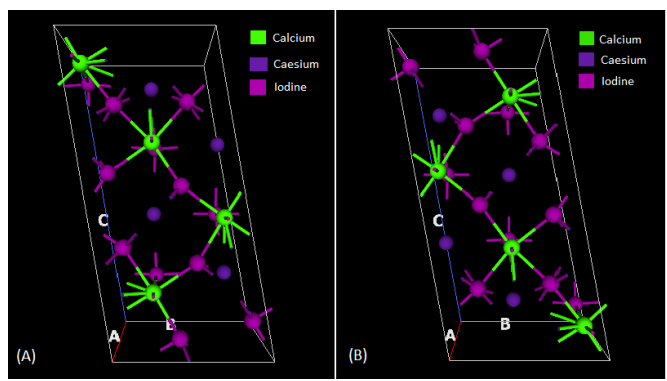

Figure 1. Unit cells of top two structures obtained from crystal structure prediction

Least energy structure shown in figure 1 (A) was chosen for substitutional doping and subsequent calculations. Doping was done by replacing 1, 2 and 3 calcium atoms in the unit cell by silicon. The resulting unit cells after geometry optimization are shown in figure 2. Cell parameters of the un-doped $CsCaI_3$ and three Si substituted structures are listed in Table 1.

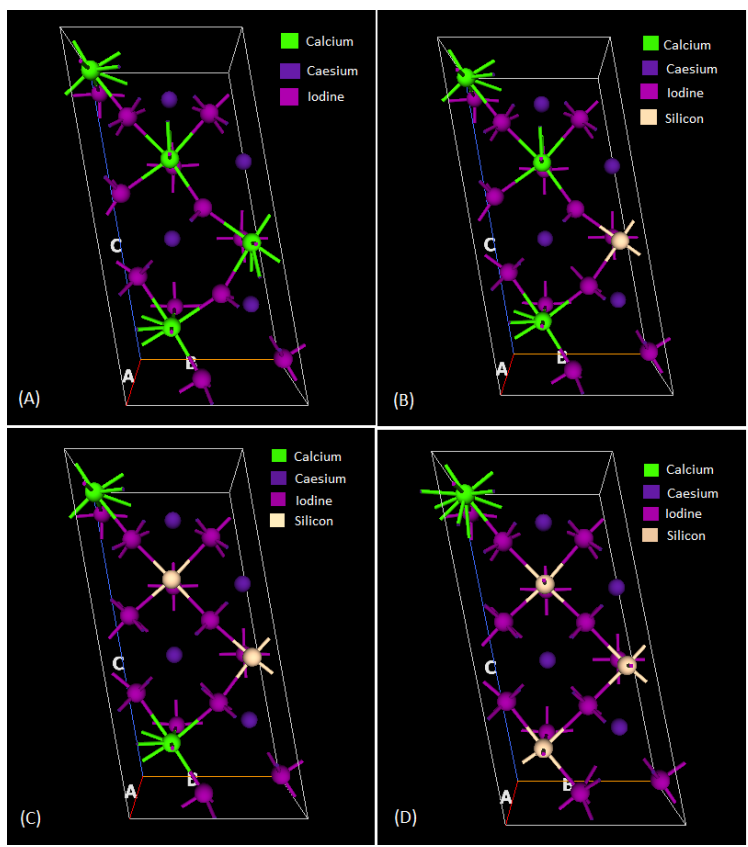

Figure 2. Unit cell structures of un-doped $CsCaI_3$ (A), $CsCa_{0.75}Si_{0.25}I_3$ (B), $CsCa_{0.5}Si_{0.5}I_3$ (C) and $CsCa_{0.25}Si_{0.75}I_3$ (D)

To observe the effect of change of cation in the structure, we calculated the bond length and bond angles of both apical and equilateral bonds of Ca-I and Si-I octahedrals. We observed that in all four structures, the octahedrals were not symmetrical. In the un-doped structure, the equatorial bond lengths varied from 3.075Å to 3.176Å while apical bond lengths ranged from 3.117Å to 3.143Å.

| Material | A (Å) | B (Å) | C (Å) | A (º) | B (º) | Γ (º) | Volume (Å³) |
|---|---|---|---|---|---|---|---|
| $CsCaI_3$ | 6.2312 | 8.5189 | 17.9981 | 99.9358 | 90.0030 | 90.0007 | 941.0707 |
| $CsCa_{0.75}Si_{0.25}I_3$ | 6.1249 | 8.4684 | 17.8388 | 100.0280 | 89.9858 | 90.0764 | 911.1429 |
| $CsCa_{0.5}Si_{0.5}I_3$ | 6.0116 | 8.3793 | 17.6706 | 99.9936 | 89.9948 | 89.9357 | 876.6217 |
| CsCa$_{0.25}$Si$_{0.75}$I$_3$ | 5.8830 | 8.3231 | 17.5877 | 101.1640 | 89.8189 | 89.8348 | 844.8788 |

Table 1. Lattice parameters of un-doped CsCaI$_3$, Single Si substituted CsCa$_{0.75}$Si$_{0.25}$I$_3$, Double Si substituted CsCa$_{0.5}$Si$_{0.5}$I$_3$ and Triple Si substituted CsCa$_{0.25}$Si$_{0.75}$I$_3$.

This variation increased when Ca was substituted by Si. In 1 atom substitution, the equatorial bond lengths and apical bond lengths were observed to vary from 2.919Å to 3.243Å and from 2.825Å to 3.062Å, respectively. This reduction is caused by the smaller Si cation. It is observed that the unit cell volume reduced with increased amount of Si atoms (shown in table 1). The bond angles of Cs-Si-Ca bonds or sides of the perovskite cube also varied with silicon substitution. Un-doped CsCaI$_3$ had lower bond angles indicating higher order of distortion in the perovskite structure. The distortion reduced as number of silicon substitution increased. Un-doped structure had a bond angle range of 154.6° to 175.2° and 164.8° to 176.4° in the equatorial and apical direction, respectively. Further, these angles were reduced to a smaller range of 168°-178.1° and 166.4°-178° in the equatorial and apical directions, respectively.

To ascertain the stability of the perovskite phase of the studied materials, we calculated the Goldschmidt tolerance factor t as given below,

$$t = \frac{(r_A + r_X)}{\sqrt{2}(r_B + r_X)} \quad \ldots\ldots\ldots\ldots\ldots (1)$$

A material in the range of tolerance factor 0.81−1.11 exists in the perovskite structure, and out of this range alternative structures can form, such as orthorhombic, hexagonal, etc. [62, 63]. The tolerance factors calculated for our structures are listed in table 2 and observed in the range of 0.8 to 1.1, confirming the formation of perovskite structure.

| Material | Tolerance factor |
|---|---|
| CsCaI$_3$ | 0.873 |
| CsCa$_{0.75}$Si$_{0.25}$I$_3$ | 0.926 |
| CsCa$_{0.5}$Si$_{0.5}$I$_3$ | 0.987 |
| CsCa$_{0.25}$Si$_{0.75}$I$_3$ | 1.053 |

Table 2. Tolerance factors of the different structures used in this study

With the increasing number of silicon atom substitutions, the tolerance factor approached the value of 1.1 which corresponds to the cubic perovskite structure as seen in the bond angle measurements mentioned earlier.

To confirm the thermodynamic stability of the materials, we calculated the decomposition energies of the most probable decomposition paths. The decomposition enthalpy of our mixed structures can be calculated by the relation,

$$\Delta H = E(CsCa_{1-n}Si_nI_3) - [E(CsI) + (1-n)E(CaI_2) + n\{E(Si) + E(I_2)\}] \dots\dots\dots\dots$$

(2)

The calculated enthalpies per unit cell of our materials are listed in table 3.

| Material | Formation Enthalpy |
|---|---|
| $CsCaI_3$ | -1.795 eV/u.c. |
| $CsCa_{0.75}Si_{0.25}I_3$ | -1.872 eV/u.c. |
| $CsCa_{0.5}Si_{0.5}I_3$ | -4.0523 eV/u.c. |
| $CsCa_{0.25}Si_{0.75}I_3$ | -5.3698 eV/u.c. |

Table 3. Formation enthalpy of the different structures used in this study

Calculated negative formation enthalpies for all four structures indicates their thermodynamic stability. Further increasing negative value with increasing Si substitution suggests improved stability of these structures.

To understand the effect of Si substitution on the band gap and electronic properties of the material, we calculated the band structures of these materials. DFT calculations are known to either underestimate or overestimate the band gaps of semiconductors. Various corrections have been suggested and new methods are proposed. Table 4 lists the band gaps calculated by different methods. Band-gap values of $CsCaI_3$ by GGA-PBE D3 method is 3.73 eV. This is an indirect band gap at Z point. It also has direct band gap of 3.77 eV between Z and X symmetry points. This is in good agreement with reported value by Debmalya Ray et al., [35] and Sabine Körbel et al., [36] using PBE-D3 method. Marina R Filip et al.,

[37] reported a band gap value of nearly 3.5 eV calculated using LDA method which is known to further underestimate the band gap value. Band gap calculated by DFT-1/2 method is 5.70 eV (Direct) and 5.62 eV (Indirect). This over estimated value is comparable to the band gap calculated by Byungkyun Kang et al., [38] using PBE0 functional. Meta GGA functional calculations gave a band gap of 4.76 eV for $CsCaI_3$ which is comparable to the results from HSE06 calculations [35]. FP-LAPW (full potential linearized augmented plane wave) calculations conducted by M. Tyagi et al., [59] estimated a band gap of 4.02 eV for the orthorhombic structure they have adopted. They observed that the band gap was under estimated as compared to experimental observations.

Both HSE06 and MGGA functionals have been reported to give most accurate band gap values for semiconductors.

| Material | PBE (eV) | GGA-PW (eV) | 1\2 DFT (eV) | HUBBARD-U (eV) | M-GGA (eV) | PBE+SOC (eV) |
|---|---|---|---|---|---|---|
| $CsCaI_3$ | 3.77(D) 3.73(ID) | 3.77(D) 3.73(ID) | 5.70(D) 5.62(ID) | 3.77(D) 3.73(ID) | 4.76(D) 4.72(ID) | 3.61(D) 3.54(ID) |
| $CsCa_{0.75}Si_{0.25}I_3$ | 1.642(D) | 1.657 | 2.278 | 1.642 | 1.80 | 1.54 |
| $CsCa_{0.5}Si_{0.5}I_3$ | 1.009 | 1.02 | 1.53 | 1.009 | 1.088 | 0.925 |
| $CsCa_{0.25}Si_{0.75}I_3$ | 0.651 | 0.672 | 1.126 | 0.651 | 0.639 | 0.576 |

Table 4. Band gaps calculated using different computational methods.

Figure 3 shows the calculated band structures using MGGA functional. $CsCaI_3$ has a flat bands near the VB maximum with small bending in the CB. While the non-perovskite orthorhombic phase in their calculation showed direct band gap at Γ point, our structure showed indirect band gap between Z and X points. With the increase in Si substitution, more and more bands appeared near the Fermi level reducing the band gap from 3.77eV for $CsCaI_3$ to 0.639eV for $CsCa_{0.25}Ca_{0.75}I_3$.

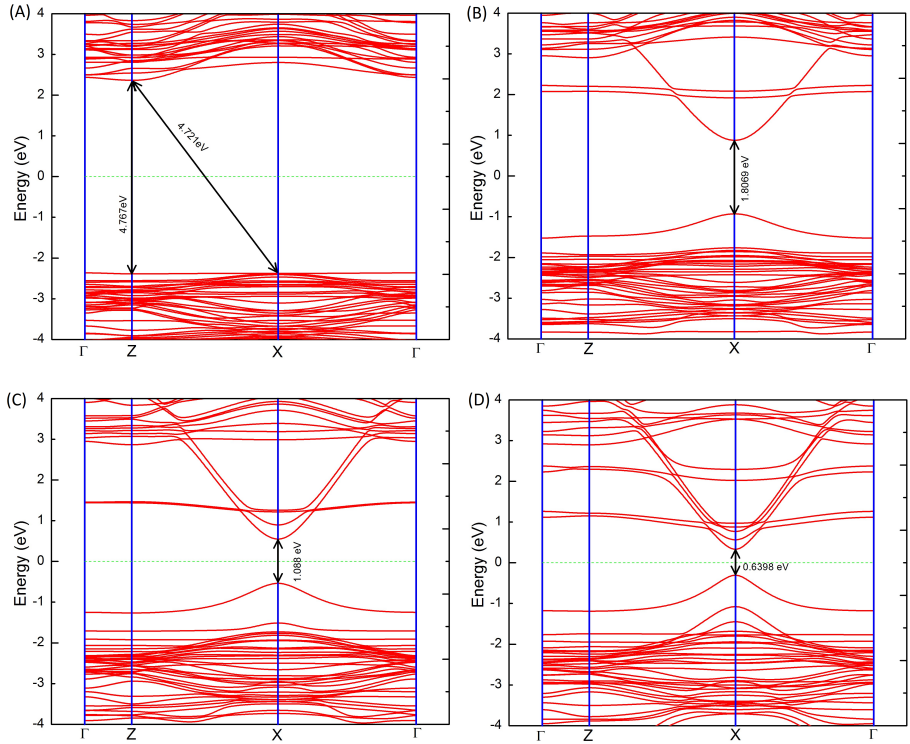

Figure 3. MGGA band structures of CsCaI$_3$ (A), CsCa$_{0.75}$Si$_{0.25}$I$_3$ (B), CsCa$_{0.5}$Si$_{0.5}$I$_3$ (C) and CsCa$_{0.25}$Si$_{0.75}$I$_3$ (D).

Substitution of Si also caused the appearance of a direct band gap at X symmetry point. Similar trend was also observed in calculations done using other DFT functionals.

The band gap values calculated using PBE, PW and Hubbard U methods were in agreement with each other for all the samples while MGGA method suggested higher values of band gap for CsCaI$_3$ and CsCa$_{0.75}$Si$_{0.25}$I$_3$. But the value predicted by MGGA functional became lower than PBE, PW and Hubbard U methods for CsCa$_{0.25}$Si$_{0.75}$I$_3$. We also calculated band gaps by considering the effect of Spin Orbit Coupling (SOC) into effect. As expected, this calculation gave slightly lower band gap values than PBE calculations because of the band splitting effect caused by SOC. Projected Density of States (PDOS) depicting the contribution of different element orbitals to the top of the valence band and bottom of the conduction band are shown in figure 4. As can be seen from the figures, in the un-doped sample, the conduction band minimum mainly consists of Iodine p states with high density and Valence band maximum consists mainly of Calcium d states with small contribution from iodine d-states. This behavior is also reported in [59] where an orthorhombic phase of CsCaI$_3$ is used for calculations. When Ca atoms are substituted with Si atoms, new states corresponding to Silicon s-orbital and Iodine p-orbitals appeared near the Fermi level in the conduction band. Similarly Si p-orbitals created new states in the valence band near the Fermi level along with slight contribution from iodine p and d orbitals. These results indicate that with Si doping the band gap can be controlled and a suitable band gap needed for PV application can be achieved by varying the doping percentage of Si in $CsCaI_3$.

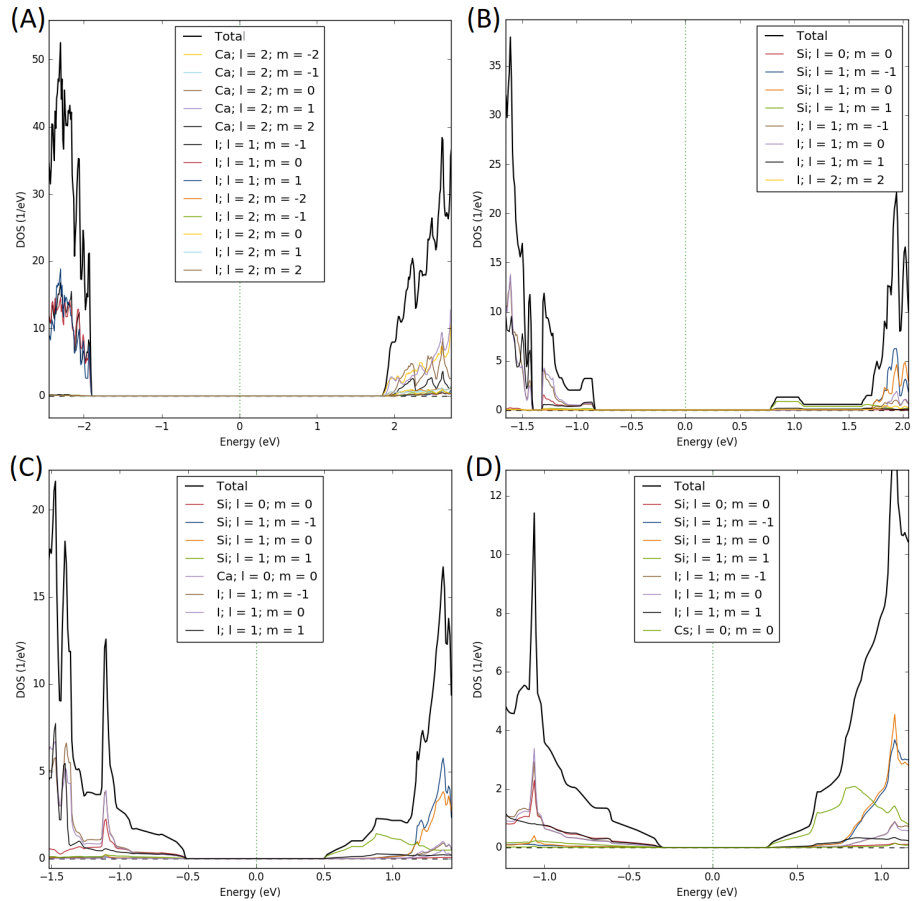

Figure 4. Projected density of states of $CsCaI_3$ (A), $CsCa_{0.75}Si_{0.25}I_3$ (B), $CsCa_{0.5}Si_{0.5}I_3$ (C) and $CsCa_{0.25}Si_{0.75}I_3$ (D)

An important property to consider when choosing materials for photo voltaic applications is its optical absorption spectrum. The material should be able to absorb most of the visible spectrum and near ultraviolet spectrum to some extent to cover the solar spectrum. Figure 5 shows the calculated optical absorption spectra for the materials under study. The plots show the XX, YY and ZZ tensors of absorption coefficients against wavelengths for all four materials under study. Since the structures studied are not symmetric in all the directions, the three tensors are not same. The total absorption can be estimated to be the average of the three tensors. Higher absorption can be achieved by growing the thin film in the direction of highest absorption. CsCaI$_3$ has sharp absorption edges with no absorption in the visible region. Upon substitution absorption peaks start appearing in the visible region as shown in figure 5. CsCa$_{0.75}$Si$_{0.25}$I$_3$ has extended absorption from ultraviolet region with sharp edge at around 400nm with smaller absorption peaks at 600nm and 720nm. When the Si substitution increased to 50%, absorption increased in the visible region with medium absorption throughout the visible spectrum. The absorption further increased when the Si substitution increased to 75% with good absorption in both ultraviolet and visible region extending to infrared region. Induction of absorption in the Si doped structures can be attributed to the appearance of Si bands in the band gap of CsCaI$_3$ as can be seen from the band structure and PDOS.

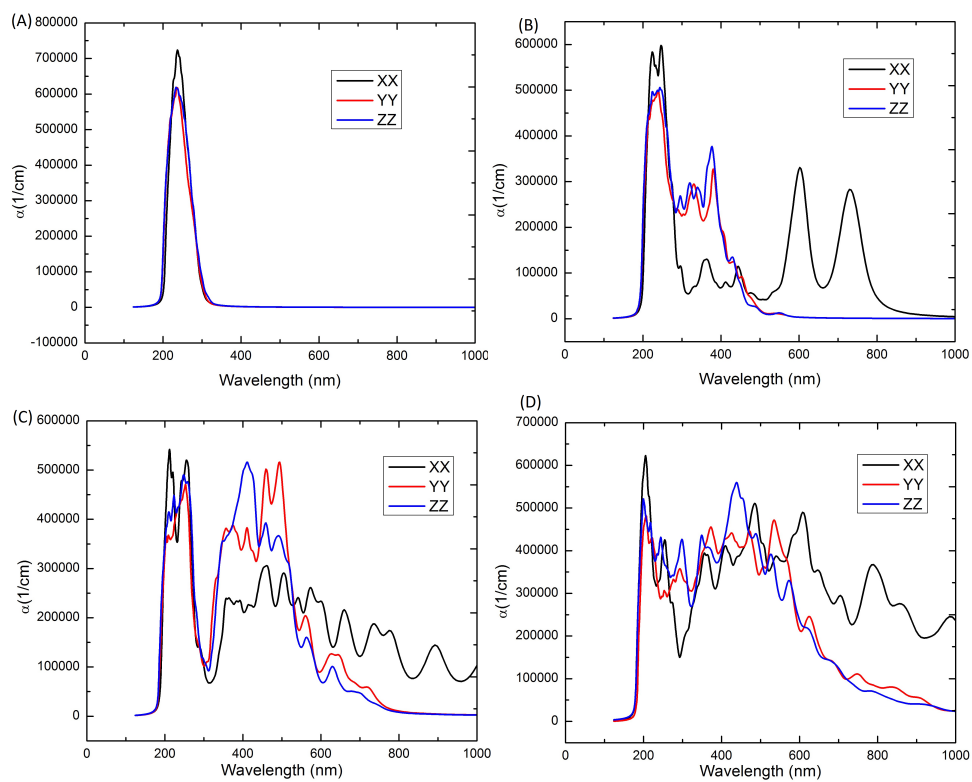

Figure 5. Absorption coefficient v/s wavelength plot of CsCaI$_3$ (A), CsCa$_{0.75}$Si$_{0.25}$I$_3$ (B), CsCa$_{0.5}$Si$_{0.5}$I$_3$ (C) and CsCa$_{0.25}$Si$_{0.75}$I$_3$ (D)

Ambipolar transport of electrons and holes in metal halide perovskites is one of the important properties which contributed to the higher power conversion efficiency of perovskite solar cells. Effective masses of electrons and holes are the indicators of high charge carrier mobility which results in ambipolar transport. Overlapping of orbitals of metal cation and halide anion also effects the mobility by providing easy conduction path. We calculated the Electron localization functions (ELF) and the effective masses of holes and electrons to analyze the effect of Si doping on the transport properties of $CsCaI_3$. Figure 6 shows the calculated ELF of out structures. As can be seen from figure 6, with increasing doping of Si, the overlapping of orbitals in Si-I and Ca-I bonds increase suggesting better carrier transport properties [64].

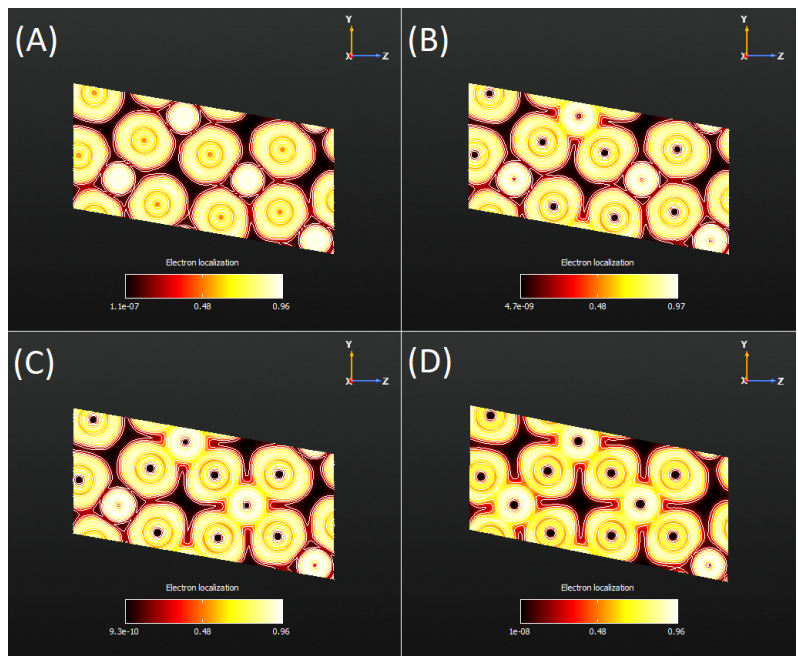

Figure 6. Electron localization functions of $CsCaI_3$ (A), $CsCa_{0.75}Si_{0.25}I_3$ (B), $CsCa_{0.5}Si_{0.5}I_3$ (C) and $CsCa_{0.25}Si_{0.75}I_3$ (D)

The effective masses charge carriers in the studied materials are listed in Table 5. For $CsCaI_3$, electrons had minimum effective mass of 0.416 $m_e$ in (101) direction at Z symmetry point and hole effective mass of 1.661 $m_e$ was minimum at X symmetry point in (100) direction.

| Material | $m_e^*$ | $m_h^*$ |
| --- | --- | --- |
| $CsCaI_3$ | 0.416 $m_e$ | 1.661 $m_e$ |
| $CsCa_{0.75}Si_{0.25}I_3$ | 0.102 $m_e$ | 0.207 $m_e$ |
| $CsCa_{0.5}Si_{0.5}I_3$ | 0.070 $m_e$ | 0.116 $m_e$ |
| $CsCa_{0.25}Si_{0.75}I_3$ | 0.043 $m_e$ | 0.056 $m_e$ |

Table 5. Calculated charge carrier effective masses of materials under study.

These values are comparable to the effective masses reported in [35] and [36]. It can be observed that with Si doping, the effective masses of both electrons and holes reduced to very low values. In case of 75% Si doping, the electron effective mass reduced to 4.3% of the electron mass and hole effective mass reached 5.6% of the electron mass. These low values indicate high mobility of electrons and holes. These effective masses are far lower than the reported effective masses in MAPbI$_3$ [39].

## 4. Conclusion

Compositional engineering to obtain desirable properties in materials is becoming more and more desirable due to the advent in both theoretical and experimental material science. In this work, we studied the effect of Si doping on the structural, electronic and optical properties of CsCaI$_3$ through First Principles calculations. From our calculations, we predict that CsCaI$_3$ can form stable perovskite structure. It's also observed that after substitutional doping of Si in CsCaI3, the material still stays in perovskite form. Band structure studies showed that with Si doping, the band gap can be varied from 4.76eV for undoped CsCaI$_3$ to 0.639eV for 75% Si doped CsCaI$_3$. The optical absorption spectra of the materials showed that Si doping can induce light absorption in the visible region in CsCaI$_3$. ELF and effective mass calculations showed that Si doping can improve the electronic conduction in this material to meet the requirements of PV applications. We hope that the results from this study can inspire further investigation into this material both theoretically and experimentally. Further study has to be conducted to study the effects of temperature on the stability of these materials and charge conduction in device level. The moisture sensitivity of these doped materials have to be investigated to ascertain the applicability as solar absorbers.

**Declaration of interest**

The authors report no conflicts of interest. The authors alone are responsible for the content and writing of this article.

**Acknowledgement**

The authors would like to thank Director, NIT Kurukshetra, Department of Physics and all the staff members of Department of Physics for their support. Authors also thank CSIR-CSIO Chandigarh for the project environment provided. Dr Parveen Kumar thanks Department of Science and Technology for

INSPIRE faculty award. Mr. Vikas Latiyan and Mr. Praveen Lakhera are acknowledged for their support.

Molecules in Hybrid Perovskite CH3NH3PbI3. Nano Lett. 2015, 15, 248-253.